\documentclass[usenatbib]{mnras}
\usepackage{aas_macros}
\usepackage{amsfonts}
\usepackage{amsmath}
\usepackage{amssymb}
\usepackage{graphicx}
\usepackage[dvipsnames]{xcolor}
\hypersetup{colorlinks=true,allcolors=teal}
\usepackage{microtype}

\newcommand{\Mh}{\ensuremath{h^{-1}M_{\odot}}}
\newcommand{\Mpch}{\ensuremath{h^{-1}{\rm Mpc}}}

\newcommand{\NJACK}{\ensuremath{25}}

\newcommand{\eqn}[1]{equation~\eqref{#1}}

\newcommand{\be}{\begin{equation}}
\newcommand{\ee}{\end{equation}}

\title[Tidal dependence of SDSS clustering]
      {The dependence of galaxy clustering on tidal environment in the Sloan Digital Sky Survey} 
\date{draft}

\author[Paranjape, Hahn \& Sheth]{
Aseem Paranjape$^{1}$\thanks{E-mail: aseem@iucaa.in}, 
Oliver Hahn$^{2}$\thanks{E-mail: oliver.hahn@oca.eu} 
\& Ravi K. Sheth$^{3,4}$\thanks{E-mail: shethrk@physics.upenn.edu}
\\  
 $^1$ Inter-University Centre for Astronomy \& Astrophysics,
      Ganeshkhind, Post Bag 4, Pune 411007, India\\
 $^2$ Laboratoire Lagrange, Universit\'e C\^ote d'Azur, Observatoire de la C\^ote d'Azur, CNRS, 
      Blvd de l'Observatoire,\\\hskip0.15in CS 34229, 06304 Nice cedex 4, France\\
 $^3$ Center for Particle Cosmology, University of Pennsylvania, 
      209 S. 33rd St., Philadelphia, PA 19104, USA\\
 $^4$ The Abdus Salam International Center for Theoretical Physics,
      Strada Costiera, 11, Trieste 34151, Italy}

\begin{document}

\label{firstpage}
\pagerange{\pageref{firstpage}--\pageref{lastpage}}

\maketitle 

\begin{abstract}
\noindent
The influence of the Cosmic Web on galaxy formation and evolution is of great observational and theoretical interest. 
We investigate whether the Cosmic Web leaves an imprint in the spatial clustering of galaxies in the Sloan Digital Sky Survey (SDSS), using the group catalog of Yang et al. and tidal field estimates at $\sim2\Mpch$ scales from the Mass-Tides-Velocity data set of Wang et al. 
We use the \emph{tidal anisotropy} $\alpha$ (Paranjape et al.) to characterise the tidal environment of groups,
and measure the redshift-space 2-point correlation function (2pcf) of group positions and the luminosity- and colour-dependent clustering of group galaxies using samples segregated by $\alpha$. 
We find that all the 2pcf measurements depend strongly on $\alpha$, with factors of $\sim20$ between the large-scale 2pcf of objects in the most and least isotropic environments. 
To test whether these strong trends imply `beyond halo mass' effects for galaxy evolution, we compare our results with corresponding 2pcf measurements in mock catalogs constructed using a halo occupation distribution that only uses halo mass as an input. 
We find that this prescription qualitatively reproduces \emph{all} observed trends, and also quantitatively matches many of the observed results. 
Although there are some statistically significant differences between our `halo mass only' mocks and the data -- in the most and least isotropic environments -- which deserve further investigation, our results suggest that if the tidal environment induces additional effects on galaxy properties other than those inherited from their host halos, then these must be weak. 
\end{abstract}

\begin{keywords}
cosmology: large-scale structure of the Universe -- galaxies: evolution
\end{keywords}

\section{Introduction}
\label{sec:intro}
\noindent
The most striking aspect of the spatial distribution of galaxies in the nearby Universe is the network of filaments connecting large groups and clusters and surrounding spatially extended underdense voids, known as the Cosmic Web \citep{zeldovich70,bbs85,bkp96,aCvdWj10}. The observational inference of the existence of the Cosmic Web relies on the galaxies \citep{hg86,gh89,colless+01} and gas \citep{akamatsu+17,tanimura+18,dGchp18} that light it up. From the physical point of view, though, it is perhaps more interesting to ask whether and how the Cosmic Web influences the formation and evolution of these galaxies and the inter-galactic medium. Hierarchical models of structure formation predict that the formation history and late time properties of collapsed dark matter haloes are, in fact, correlated with their location in the Cosmic Web \citep{ct96a,acc06,hahn+07,hahn+07b,codis+12,fRcp14}, with haloes in filaments experiencing very different tidal forces and mass inflow rates than those in more isotropic environments \citep{bprg17}. Since galaxies are believed to form in the gravitational potential wells of these haloes \citep{wr78}, one expects a similar influence of the Cosmic Web on the amount and thermal properties of gas being fed into newly forming galaxies \citep{dubois+14,rgbp17}.

There are indeed direct observational indications of such an influence in targeted small volumes \citep[see, e.g.,][]{kodama+01,mrp12,coppin+12,darvish+14}. There is also considerable literature on statistical analyses of samples controlled by local galaxy overdensity that show interesting trends in stellar masses, star formation rates, gas fractions, etc. with distance from the spines of filaments \citep{alpaslan+16,ktt17,chen+17,malavasi+17,laigle+18,kraljic+18,odekon+18}. 

Most statistical analyses of the \emph{clustering} of large samples of galaxies, however, are consistent with a simpler picture of galaxy evolution. This states that halo mass is the primary variable relevant for the physics of galaxy formation and evolution, and that all other observed environmental trends \citep[including the `classical' trends of colour and morphology noted by, e.g.,][]{ms77,bo78,dressler80} can be largely explained as being `inherited' from the spatial locations of the haloes themselves, along with the dichotomy of central and satellite populations \citep{cs02,zehavi+05,zheng+07,ss09,zehavi+11,coupon+15}. In other words, in this picture, the physics of galaxy evolution need not explicitly depend on, e.g., whether or not the galaxy happens to reside in a cosmic filament:  the only environment which matters is the mass of its host halo.  In the context of the discussion above, it is interesting to note that most Cosmic Web-related observational analyses to date do not or cannot control for halo mass \citep[although see][]{poudel+17}. 
Overall, it is therefore very interesting to look for potential `smoking gun' observational signatures, in large samples of galaxies, of the influence of the Cosmic Web on galaxy evolution beyond what is accounted for by halo mass alone. 

There has been considerable recent interest in predicting and detecting such `beyond halo mass' effects, broadly referred to as `assembly bias' in the literature \citep{st04,gsw05,wechsler+06,hearin+15a}, both at low \citep{tojeiro+17} and high redshifts \citep{hj17} and across a range of halo masses \citep{lin+16,miyatake+16,montero-dorta+17}. As yet, there is no consensus on a robust determination of assembly bias in observed samples, primarily due to difficulties with constructing sufficiently pure samples \citep{twcm17,zu+17}. In this work, we approach the problem from a somewhat different point of view.

Theoretical studies have established a strong correlation between the local tidal environment and the abundance \citep{sams06} as well as the large-scale spatial clustering \citep{hahn+09,yang+17,phs18} of dark matter haloes. Specifically, halo clustering strength is seen to be a strong function of the \emph{tidal anisotropy} of the immediate halo environment (see below), with clustering strength increasing from isotropic to anisotropic environments at any fixed halo mass \citep{phs18}. The tidal environment (e.g., whether or not a halo is in a filament) also plays a significant role in governing the mass accretion history of haloes at fixed present-epoch mass \citep{hahn+09,hbv16,bprg17} and has been shown by \citet{phs18} to be a primary suspect for explaining mass dependent trends in the correlation between halo age/concentration and large scale clustering. As such, we expect the tidal anisotropy around haloes/galaxies to be a simple and useful diagnostic of `beyond halo mass' effects.
The `halo mass only' approach to galaxy formation and evolution, for example, would predict that galaxy clustering should inherit a strong dependence on tidal anisotropy when averaged over halo mass, but that the luminosity- and colour-dependence of clustering should not show any dependence on tidal environment at fixed halo mass. 

Testing this requires, of course, estimates of the tidal field in the volume of a galaxy survey. Such measurements are now becoming available \citep{wmyv12,jlw15}. In this work, we will use estimates of the tidal field (specifically, the eigenvalues of the tidal tensor evaluated on a spatial grid) in a sub-volume of the Sloan Digital Sky Survey provided by \citet[][henceforth, W12]{wmyv12} along with group membership information from \citet[][henceforth, Y07]{yang+07}. We will focus on the dependence of redshift-space clustering on the tidal environment, as well as the effect of the latter on the luminosity- and colour-dependence of clustering. Ideally, one would want to perform these clustering measurements at fixed halo mass; however, the observational estimates of halo masses in the Y07 group catalog have large uncertainties. In order to test the `halo mass only' model of galaxy evolution, we will therefore compare the clustering measurements in the data with corresponding measurements in mock catalogs constructed using a halo occupation distribution (HOD) that only uses halo mass as an input.  In this respect, our approach is similar to that of \citet{as07} who showed that `halo mass only' explains almost all the dependence of galaxy clustering on the overdensity -- rather than tidal field -- of the environment.

The paper is organised as follows. We describe our data sets and mock catalogs in section~\ref{sec:data}. We present our clustering results and the comparison with the mocks in section~\ref{sec:2pcf}. We discuss these results in section~\ref{sec:discuss} and conclude in section~\ref{sec:conclude}. The Appendix gives technical details relevant to some of our analysis. Unless stated otherwise, we will assume a flat Lambda-cold dark matter ($\Lambda$CDM) cosmology with total matter density parameter $\Omega_{\rm m}=0.276$, baryonic matter density $\Omega_{\rm b}=0.045$, Hubble constant $H_0=100h\,{\rm kms}^{-1}{\rm Mpc}^{-1}$ with $h=0.7$, primordial scalar spectral index $n_{\rm s}=0.961$ and r.m.s. linear fluctuations in spheres of radius $8\Mpch$, $\sigma_8=0.811$, with a transfer function generated by the code \textsc{camb} \citep{camb}.\footnote{http://camb.info}

\section{Data set and mock catalogs}
\label{sec:data}
\noindent
In this section we describe the data sets we use in this work, along with the mock catalogs we generated to compare with the data. All data sets are based on data release 7 \citep[DR7;][]{abazajian+09} of the Sloan Digital Sky Survey\footnote{http://www.sdss.org} \citep[SDSS;][]{york+00}.

\subsection{SDSS-MTV data}
\label{subsec:sdss}
\noindent
From the Mass-Tides-Velocity (MTV) data provided by W12, we use 7 of 8 files: filling fraction $F$, 3 tidal tensor eigenvalues and 3 velocity field components (we don't use the density field in this work). The fields are defined on a rectangular subset of a $1024^3$ grid, with grid spacing $0.709\Mpch$. These are based on a subset of the Y07 group catalog restricted to DR7 data from the Northern Galactic Cap. We impose cuts below restricting the Y07 catalog to objects in this volume, having comoving $X,Y$ and $Z$ positions (in \Mpch) in the ranges $(-351.0,-1.5)$, $(-329.7,302.0)$ and $(-22.0,330.3)$, respectively. Here $X,Y,Z$ are computed using the observed redshifts and $\Omega_{\rm m}=0.258$ as done by W12 (note that Y07 used a different value for $\Omega_{\rm m}$ to define luminosities). 

The tidal tensor eigenvalues $\{\lambda_j\}_{j=1}^3$ reported by W12 were defined with a Gaussian smoothing scale enclosing a mass $10^{13}\Mh$ or a radius of $\sim2\Mpch$ (see their equation~8). These must be non-dimensionalised by dividing each by $4\pi G\bar\rho a^2\propto(1+z)$. We approximate this redshift-dependent correction by its value at $z_{\rm mid}\equiv(z_{\rm min}+z_{\rm max})/2$, which introduces a $\lesssim5\%$ systematic error in each $\lambda_j$. These eigenvalues are then used to construct the tidal anisotropy field $\alpha$ defined as
\begin{align}
\alpha &\equiv (1+\delta)^{-1}\sqrt{q^2}\,,
\label{eq:alphadef}
\end{align}
where the density contrast $\delta$ and tidal shear $q^2$ are defined in terms of the non-dimensionalised eigenvalues as
\begin{align}
\delta &\equiv \lambda_1 + \lambda_2 + \lambda_3\,, \notag\\
q^2 &\equiv \frac12\left[(\lambda_1-\lambda_2)^2 + (\lambda_2-\lambda_3)^2 + (\lambda_1-\lambda_3)^2\right]\,.
\label{eq:deltaq2def}
\end{align}
By construction, $\alpha$ involves a ratio of eigenvalues and is hence less affected by the systematic error mentioned above. 

\citet{phs18} demonstrated that $\alpha$ measured at a smoothing scale a few times larger than the halo radius correlates strongly with large-scale halo bias over a wide range of halo masses, and that this correlation is in fact stronger than the one between halo bias and halo-centric dark matter overdensity defined at the same scale as $\alpha$. Moreover, the probability distribution of $\alpha$ distinguishes quite sharply between traditionally defined node- and filament-like environments. They concluded that $\alpha$ is therefore a very useful diagnostic of halo-scale tidal effects beyond those captured by local overdensity. In the present work, we only have access to $\alpha$ defined at a fixed smoothing scale $\sim2\Mpch$; we have checked, however, that this quantity also has the same qualitative behaviour in simulations as the one studied in \citet{phs18}. We will comment on some interesting differences between the two choices of smoothing scales later.

It is also worth emphasizing, that although the tidal tensor eigenvalues were inferred by W12 using group locations from the Y07 catalog (which are a biased tracer of the dark matter distribution), the analysis of W12 removes most of this dependence through the simple model $\delta_{\rm grp} = b \delta_{\rm m}$ (see the discussion around equation~6 of W12). At leading order, therefore, we expect the eigenvalues to refer to the \emph{unbiased} dark matter field.

\begin{figure}
\centering
\includegraphics[width=0.485\textwidth]{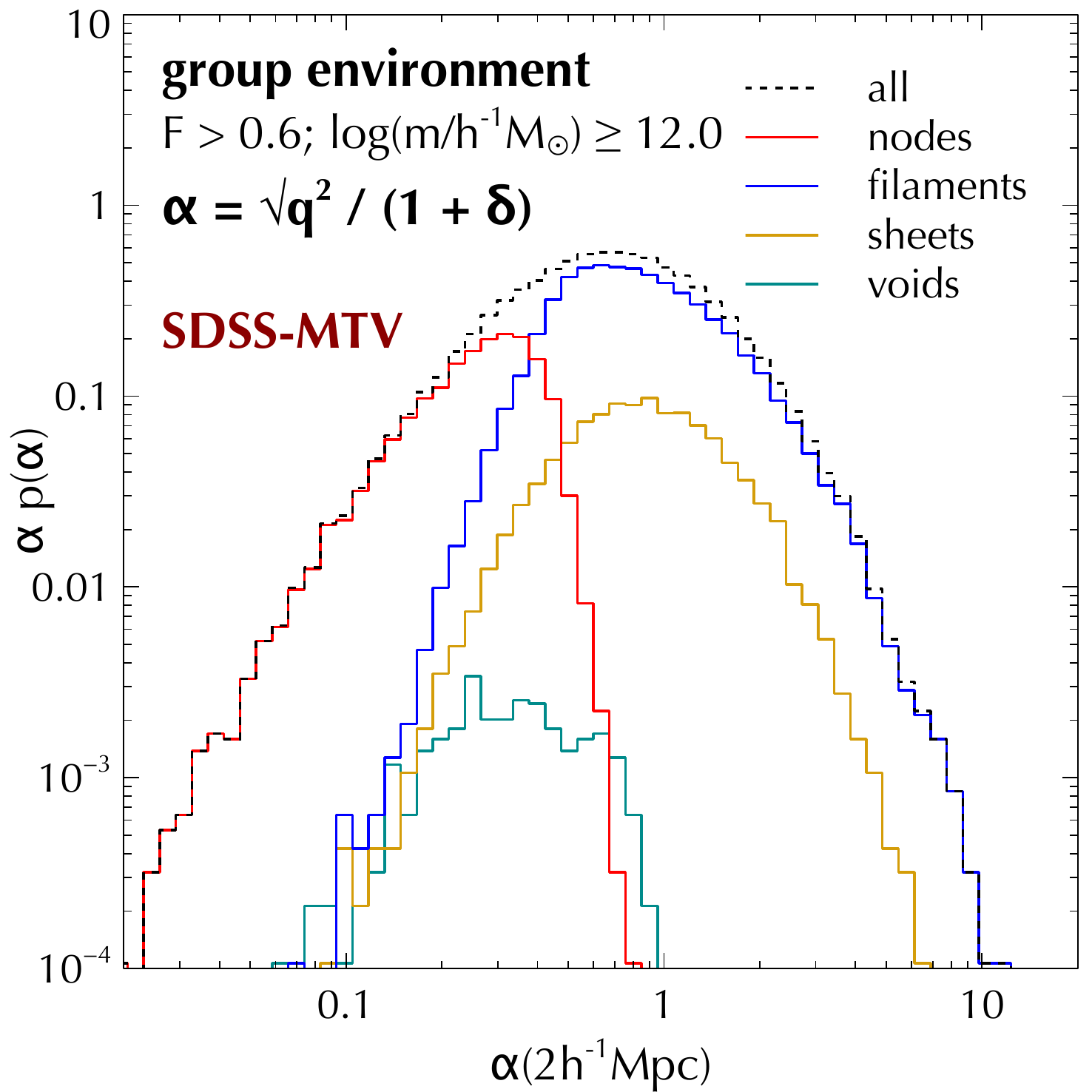}
\caption{Distribution of tidal anisotropy $\alpha$ (equation~\ref{eq:alphadef}) split by traditional web environment (with nodes, filaments, sheets and voids corresponding to $3$, $2$, $1$ and $0$ positive eigenvalues of the tidal tensor, respectively) at group locations in the SDSS-MTV region. The measurements used Gaussian smoothing with a filter of radius $2\Mpch$.}
\label{fig:alphahist}
\end{figure}

Figure~\ref{fig:alphahist} shows the distribution of $\alpha$ split by traditional web environment (defined by number of positive eigenvalues of the tidal tensor) at group locations, with each group being assigned a value of $\alpha$ as described in Appendix~\ref{subapp:rsdcorr}. Notice that the distribution is dominated by node-like environments at small $\alpha$ and filament-like environments at large $\alpha$, with the transition occurring around $\alpha\sim0.4$. This is qualitatively consistent with the findings of \citet{phs18}.

\subsection{Group positions and group galaxies from Y07}
\label{subsec:y07}
\noindent
In the following, we present clustering measurements of group positions (with groups split by tidal environment), as well as galaxy positions (with galaxies split by luminosity, colour and tidal environment). The data samples for both measurements are based on the SDSS-DR7 group catalog of Y07, which is built using the halo-based group finder described in \cite{yang+05} to identify groups in the New York University Value Added Galaxy Catalog \citep[NYU-VAGC;][]{blanton+05}, based on the SDSS. The samples are described below. In both cases, we use the $f_{\rm edge}$ information to select groups and group galaxies away from the boundaries (keeping objects with $f_{\rm edge}>0.6$). Further, we only keep objects in the SDSS-MTV volume where the MTV filling fraction $F$ satisfies $F>0.6$ as suggested by W12. Throughout, we will only consider galaxies with spectroscopic redshifts (`sample I' of Y07).

\subsubsection{Y07 subsample for group positions}
\label{subsec:y07:grp}
\noindent
For the group clustering analysis, we use the luminosity weighted group position from the Y07 catalog (given as RA, Dec, redshift). We further restrict the sample to groups having luminosity ranked halo mass $\log_{10}(m/\Mh) > 12.0$, with redshifts in the range $z_{\rm min} = 0.01 \leq z \leq 0.12 = z_{\rm max}$. This corresponds to the same criteria that were used by W12 to construct the sample used in estimating the tidal field in the SDSS-MTV volume. This leaves us with an approximately volume limited sample of $80,700$ groups in a comoving volume $\simeq(270\Mpch)^3$.

\subsubsection{Y07 subsample for galaxy positions}
\label{subsec:y07:gal}
\noindent
For the galaxy clustering analysis, we construct a volume complete sample of Y07 group galaxies with $M_r\leq-20.0$.\footnote{Here $M_r\equiv M_{{}^{0.1}r}-5\log(h)$ is the Petrosian SDSS $r$-band absolute magnitude, K-corrected to $z=0.1$ and corrected for passive evolution \citep[][see also Y07]{blanton+03}. We additionally use the `model' ${}^{0.1}(g-r)$ colours for these galaxies, which we will denote by $g-r$.} This requires two changes to the halo mass and redshift cuts mentioned above. Firstly, we relax the halo mass cut to $\log(m/\Mh) \geq 11.5$ (i.e., all halo masses present in the Y07 catalog). Secondly, we restrict the galaxy redshifts to the range $0.01\leq z\leq0.1$. This results in a volume complete sample with $87,000$ galaxies having $M_r\leq-20.0$ in a comoving volume $\simeq(240\Mpch)^3$. These choices allow us to maximise the number of galaxies in the SDSS-MTV volume while still maintaining volume completeness of the sample which allows for straightforward estimations of the clustering signal.

\subsection{Mock catalogs}
\label{subsec:mocks}
\noindent
We have generated $10$ mock catalogs using the halo-mass-only HOD calibrated by \citet{zehavi+11} which provides information on galaxy luminosities, together with the prescription of \citet{ss09} for including colours, with DR7 colour-luminosity calibrations (double Gaussian fits for $g-r$ colour at fixed $M_r$) taken from \citet{pkhp15} and the satellite red fraction from \citet[][their equation 41]{pp17b}. 

\begin{figure*}
\centering
\includegraphics[width=0.495\textwidth]{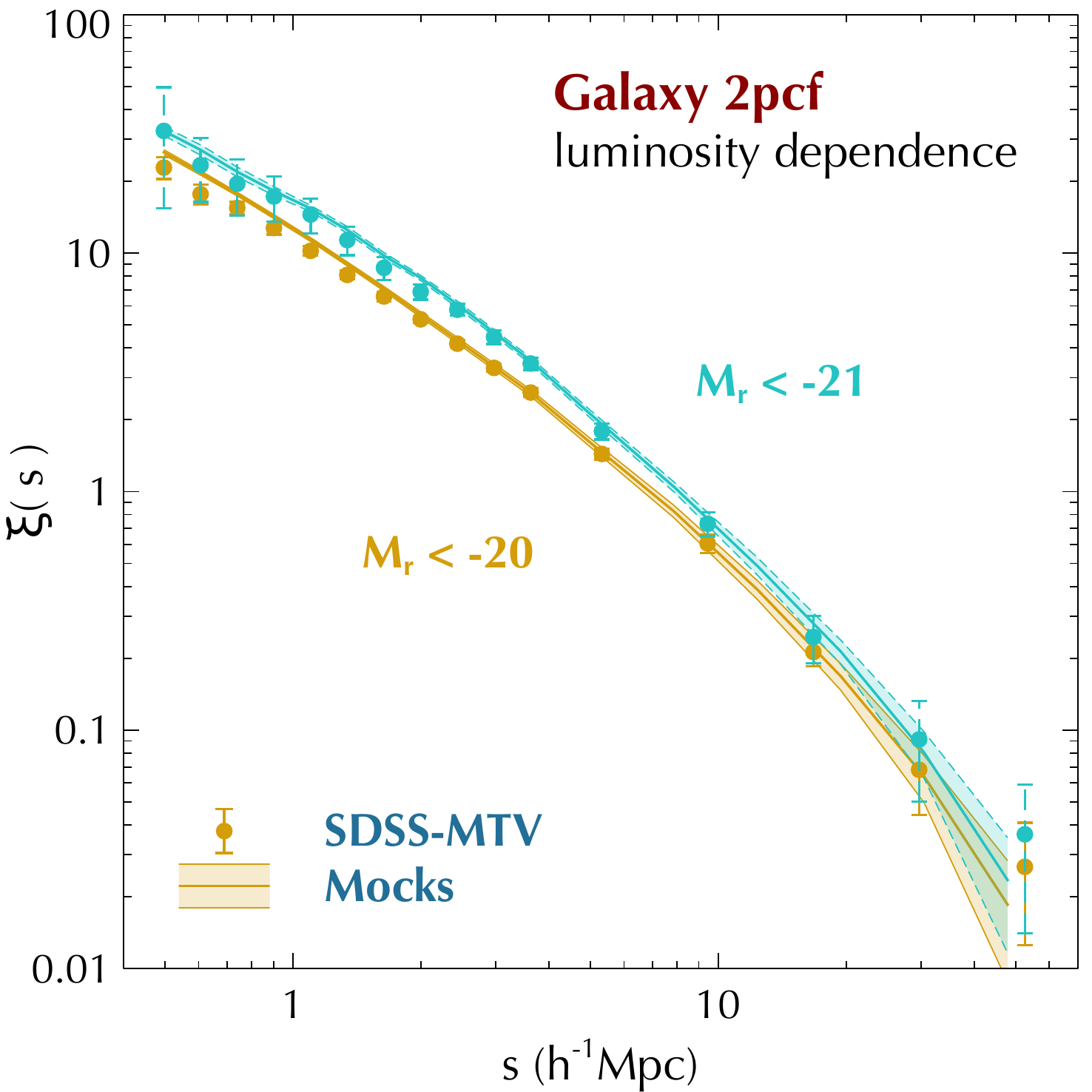}
\includegraphics[width=0.495\textwidth]{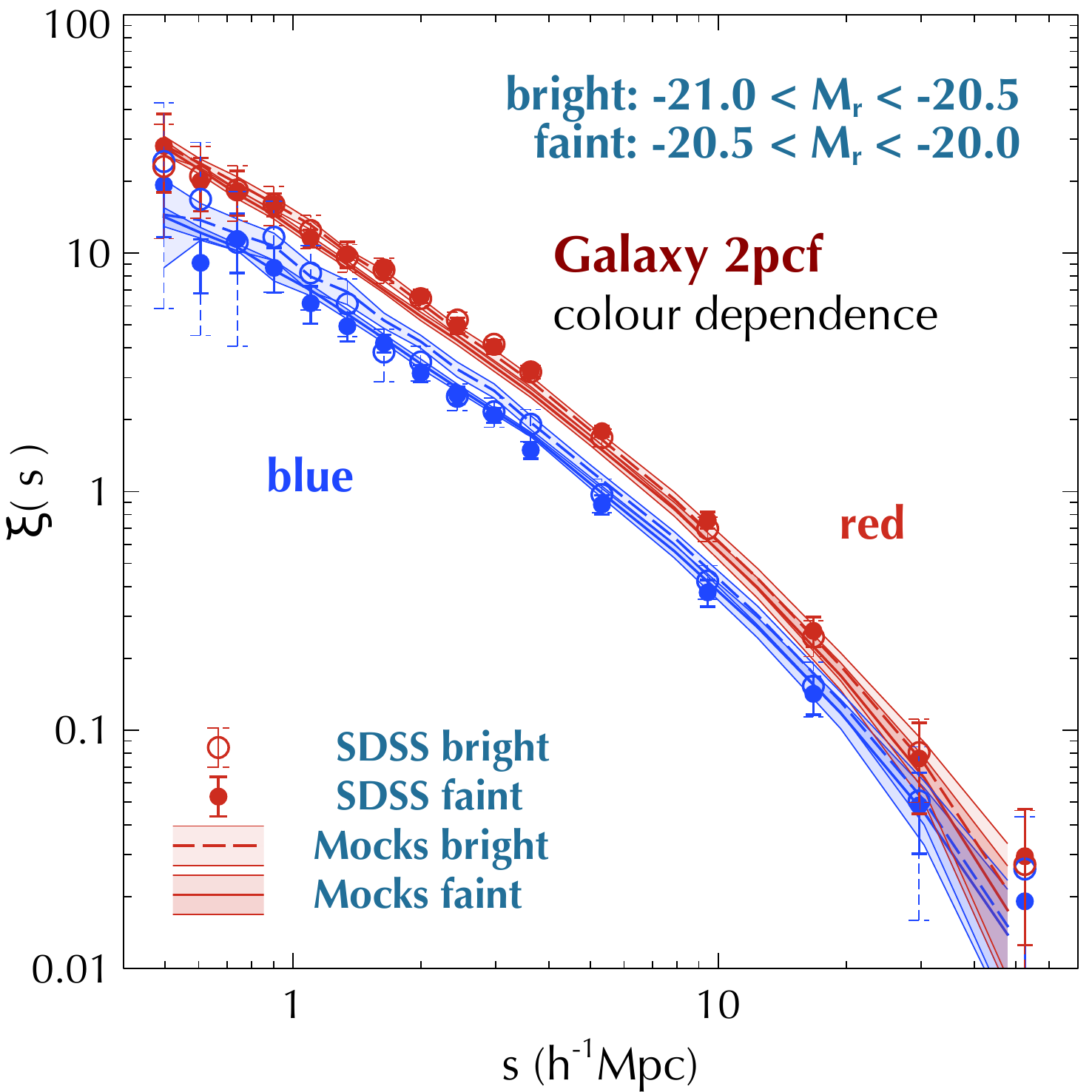}
\caption{Comparison of 2-point correlation function (2pcf) of group galaxies from the Y07 catalog in the SDSS-MTV volume (symbols with error bars) with measurements in mock catalogs (lines with error bands), both selected to be volume-complete for $M_r<-20.0$ (see text for a description of the samples).
\emph{(Left panel):} Split by $M_r$ as indicated. We see the well-known trend that clustering strength increases with luminosity. 
\emph{(Right panel):} Split by colour into `red' and `blue' according to whether the $g-r$ colour is, respectively, larger or smaller than the threshold $(g-r)_{\rm cut}=0.8-0.03(M_r+20)$. We see the well-known trend that red galaxies cluster more strongly than blue galaxies by a factor $\sim2$ in the 2pcf, at fixed luminosity.}
\label{fig:2pcf-gal-LumCol}
\end{figure*}

Our algorithm is an updated version of the one first described by \citet{sscs06} and extended by \citet{ss09}.
Central and satellite luminosities are assigned by sampling the respective conditional luminosity function determined from the HOD, with satellite number in each occupied halo also being determined by Poisson sampling the HOD. A particular galaxy's colour is assigned in two steps in our algorithm: first, a decision is made as to which Gaussian mode (red or blue) the colour should be drawn from, and the appropriate mode is then randomly sampled. Using the \citet{ss09} prescription means that both of these steps depend \emph{only} on the galaxy's luminosity and not on halo mass or any environmental variable. 

The simulations used to build these mocks are the $(300\Mpch)^3$ boxes described in \citet{phs18}, which used the cosmological parameters described in the Introduction. The matter density parameter in the simulations, $\Omega_{\rm m}=0.276$, is slightly larger than the value $\Omega_{\rm m}=0.25$ assumed by \citet{zehavi+11} for calibrating their HOD. This leads to some systematic errors in the clustering measurements that we comment on in Appendix~\ref{subapp:mockVsdata}. The dark matter particle mass in these boxes is $m_{\rm p}=1.93\times10^9\Mh$ and we use haloes with $150$ particles and higher; this allows us to achieve near-completeness for galaxies with $M_r<-20.0$. Since we have $10$ realisations of the simulation volume generated using independent initial conditions, we can generate $10$ independent mocks which we will use below.

Our algorithm places each central galaxy at the center-of-mass of its parent halo and distributes satellites around the central assuming an NFW profile \citep{nfw97} with concentration $c_{\rm 200b}=R_{\rm 200b}/r_{\rm s}$, where $R_{\rm 200b}$ and $r_{\rm s}$ are the halo radius\footnote{$R_{\rm 200b}$ is defined as the radius where the enclosed density is $200$ times the background density. The mass enclosed inside $R_{\rm 200b}$ is denoted $m_{\rm 200b}$.} and halo scale radius, respectively, as measured in the simulation.\footnote{We use the \textsc{rockstar} halo finder \citep[][http://code.google.com/p/rockstar/]{behroozi13-rockstar}, which provides information on $m_{\rm 200b}$ and $r_{\rm s}$, among many other halo properties.} Both centrals and satellites are assigned the bulk velocity of their parent halo. In addition, satellites are assigned random velocities consistent with the velocity dispersion of the NFW profile \citep[see, e.g., Appendix A3 of][]{shds01}. We ignore any velocity offsets between the galaxies and their respective parent haloes \citep{guo+15b}; we comment on this later. For further details on the algorithm, we refer the reader to section~2 of \citet{pkhp15}.

In the analysis below, we will compare clustering measurements in the mocks with those in the SDSS-MTV volume, segregating galaxies by tidal anisotropy $\alpha$. While our main comparison will be an apples-to-apples one using $\alpha$ defined using a fixed Gaussian filter with radius $R=2\Mpch$ as in W12, later we will also display mock measurements using $R$ defined as the Gaussian equivalent of $4R_{\rm 200b}$ as advocated by \citet{phs18}. The latter measurement places some restrictions on the minimum halo mass by requiring that the grid used in the tidal tensor calculations provide a fine enough sampling of the smallest $4R_{\rm 200b}$ spheres used. 
For simplicity, in the entire analysis below we will discard mock galaxies in haloes with $m_{\rm 200b}<10^{11.6}\Mh$, slightly higher than the $150$ particle limit, which corresponds to using at least $2$ grid cells in the smallest $4R_{\rm 200b}$ spheres. This is not, of course, an issue for the main $R=2\Mpch$ measurements, since this radius is very well resolved by the $512^3$ grid used by \citet{phs18}. 

\begin{figure*}
\centering
\includegraphics[width=0.9\textwidth]{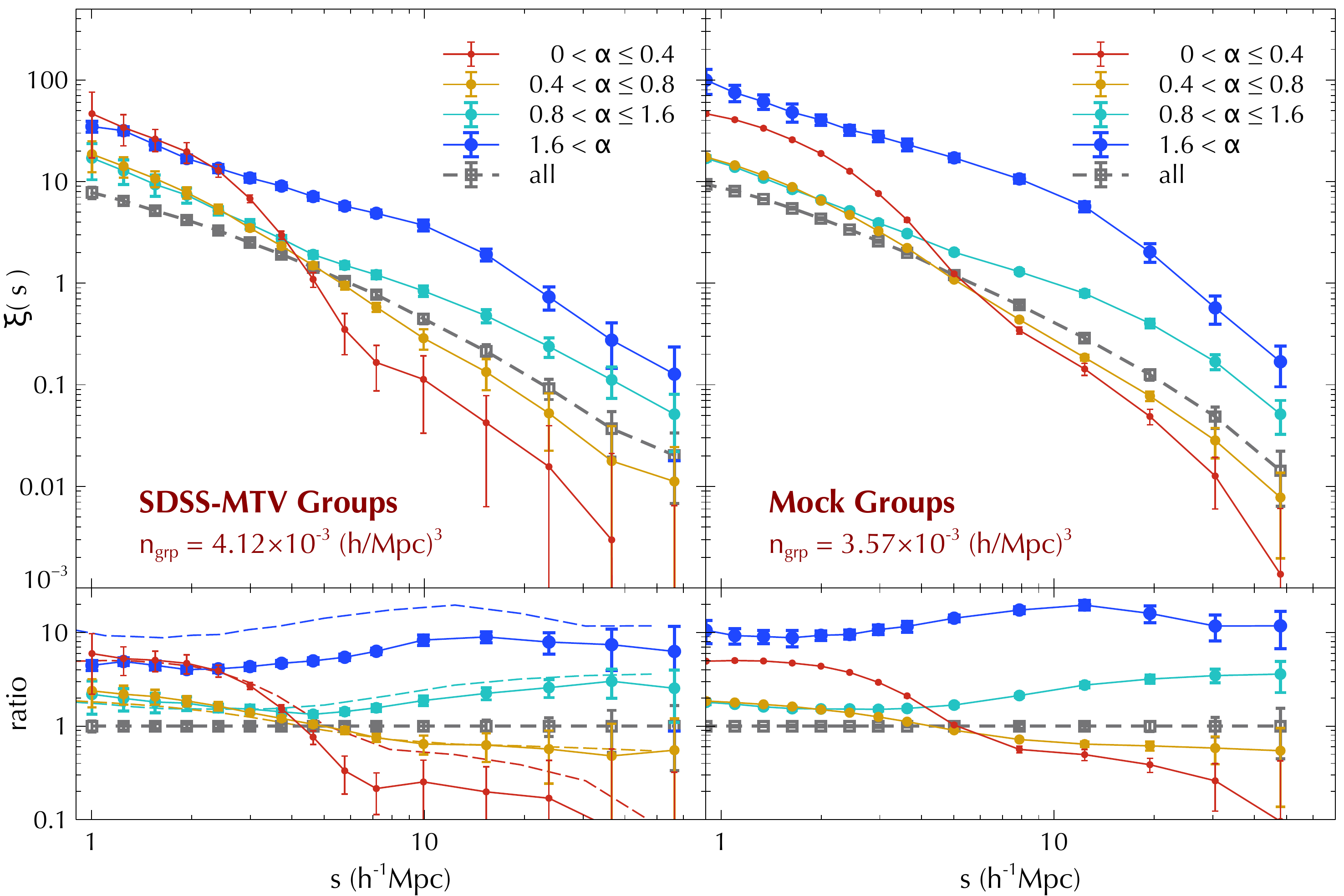}
\caption{\emph{(Top panels:)} 2-point correlation function (2pcf) for subsamples of the Y07 groups in the SDSS-MTV volume \emph{(top left panel)} and in our mock catalogs \emph{(top right panel)}. Groups were selected using $\log(m/\Mh) \geq12.0$ in each case, and were restricted to $0.01\leq z\leq0.12$ in the SDSS data and $z=0.0$ in the mocks. Results are shown for the 2pcf of the full sample (empty gray squares with dashed line) and of groups in four bins of $\alpha$ (filled red to blue circles increasing in size with $\alpha$ as indicated, with solid lines). The $\alpha$-assignment for the Y07 groups used velocity-corrected positions (Appendix~\ref{subapp:rsdcorr}) but the 2pcf used redshift-space (i.e., uncorrected) positions. \emph{(Bottom panels:)} Ratio of 2pcf in different environments to the mean trend of the full sample, colour-coded as in the top panels. Points with errors in the \emph{left} (\emph{right}) panel show results from the SDSS-MTV volume (mock catalogs). Dashed lines in the left panel show the mean mock trends from the right panel for direct comparison.}
\label{fig:2pcf-alpha}
\end{figure*}

Note that the volume in each of our boxes is about a factor $2$ larger than the effective SDSS volume used in the MTV analysis. We use the distant observer approximation to place galaxies in redshift space, and define red and blue galaxies in the same way as in the observed sample. 

\subsection{Luminosity and colour dependence of clustering}
\label{subsec:lumcol-tidal}
\noindent
Figure~\ref{fig:2pcf-gal-LumCol} shows the redshift-space 2-point correlation function (2pcf) $\xi(s)$, as a function of redshift-space separation $s$, of the Y07 group galaxies (symbols with error bars) and our mock galaxies (lines with error bands). 
We used the Landy-Szalay estimator \citep{ls93} with unweighted pair counts for the measurements in both the data as well as the mocks. For the data, we used randoms sampled according to the SDSS survey angular selection function \citep{sthh08-mangle,ht04,blanton+05}\footnote{The SDSS angular selection function produced using the \textsc{Mangle} software was downloaded from http://space.mit.edu/$\sim$molly/mangle/download/data.html. We used the Python software \textsc{pymangle} developed by E. Sheldon (https://github.com/esheldon/pymangle) which is based on C++ code developed by M. White (https://github.com/martinjameswhite/litemangle).}, further masking out the region with MTV filling factor $F<0.6$ (see section~\ref{subsec:y07}). For the mocks, we use randoms generated uniformly in our cubic simulation volumes. Here and below, error bars were estimated using \NJACK\ jackknife subsamples for the SDSS data and the standard deviation over $10$ realisations for the mocks. 

The \emph{left panel} of the Figure shows the 2pcf for two luminosity thresholds, while the \emph{right panel} shows the 2pcf split by colour (red or blue) in two luminosity bins. The red/blue classification used the colour threshold given by equation~(8) of \citet{ss09}: $(g-r)_{\rm cut}=0.8-0.03(M_r+20)$, with $M_r$ defined as above. We see the well-known trends that clustering strength increases with luminosity, and increases from blue to red galaxies at fixed luminosity. Note, however, that these trends are relatively weak, with the largest difference being a factor $\sim2$ between the 2pcf of red and blue galaxies.

Importantly, we see that the measurements in the mock catalogs are in reasonable agreement with the SDSS measurements, although the colour dependence is a bit weaker and the small-scale 2pcf is somewhat overestimated in the mocks. 
This is despite the fact that the mocks are built on simulations that have a slightly different cosmology than the one used to calibrate the HOD, as well as the fact that the mocks ignore effects such as position and velocity offsets between galaxies and their parent dark matter halo. This indicates that the resulting cumulative systematic effect is small, and we conclude that the mocks are working correctly as far as the observables they were built to reproduce are concerned. We will comment later on observables that could potentially be affected by our neglect of these additional effects.

\section{Tidal dependence of clustering}
\label{sec:2pcf}
\noindent
In this section, we present our results for the tidal dependence of clustering of groups and group galaxies in the SDSS-MTV volume, compared with corresponding measurements in our mock catalogs. We will discuss this comparison further in section~\ref{sec:discuss}.

\begin{figure*}
\centering
\includegraphics[width=0.9\textwidth]{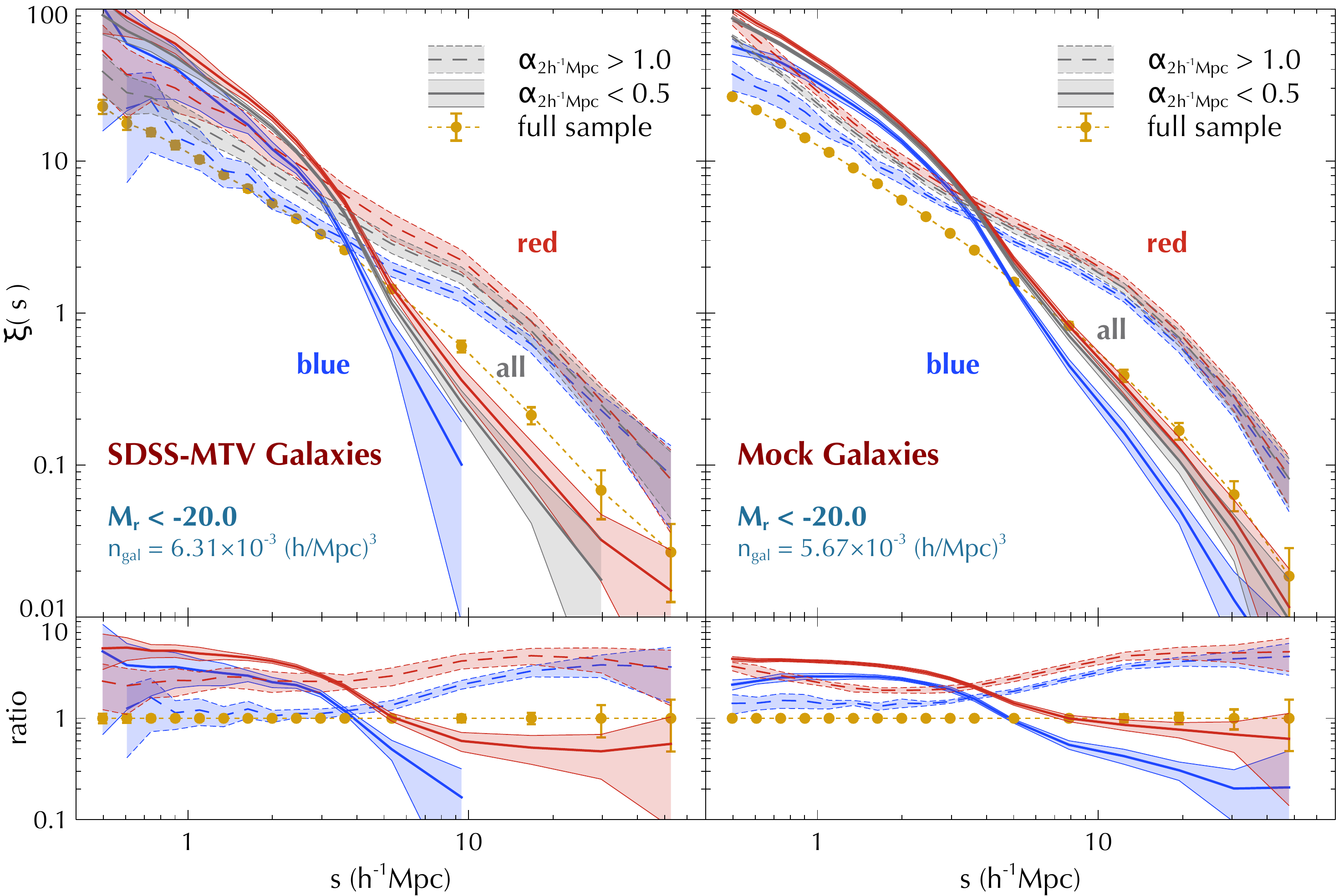}
\caption{\emph{(Top panels:)} 2-point correlation function (2pcf) of group galaxies (lines with error bands) in the SDSS-MTV volume \emph{(top left panel)} and in our mock catalogs \emph{(top right panel)}, selected to be volume-complete for $M_r<-20.0$ (the number density is indicated in the labels). We show the 2pcf for galaxies in isotropic ($\alpha<0.5$, solid lines) and anisotropic environments ($\alpha>1.0$, dashed lines). In each environment, the samples are further divided into all colours (gray), red and blue galaxies (coloured accordingly), with the red/blue split defined as in Figure~\ref{fig:2pcf-gal-LumCol}. The all-galaxy 2pcf in each case is shown for reference as the yellow symbols with error bars. As expected from the group 2pcf, we see that the galaxy clustering strength is a strong function of tidal anisotropy, with differences of a factor $\sim10$-$20$ between isotropic and anisotropic environments. \emph{(Bottom panels:)} Ratio of 2pcf of red and blue galaxies in different environments to the mean trend of the full sample, colour-coded as in the top panels and with the same line styles. 
While broadly consistent with the data, the mocks tend to underestimate the difference between red and blue galaxy clustering in both isotropic as well as anisotropic environments (c.f. Figure~\ref{fig:2pcf-gal-LumCol}). 
See text for a discussion.}
\label{fig:2pcf-gal-alpha}
\end{figure*}

Figure~\ref{fig:2pcf-alpha} shows measurements of the redshift-space $\xi(s|\alpha)$ measured for all groups and for groups in four bins of $\alpha$ as indicated, for the SDSS-MTV volume \emph{(top left panel)} and in our mocks \emph{(top right panel)}. Group locations in the SDSS-MTV volume were selected as described in section~\ref{subsec:y07:grp}, while group locations in the mocks were taken to be the centers-of-luminosity of occupied haloes with $m_{\rm 200b}\geq10^{12}\Mh$. Values of $\alpha$ were assigned to Y07 groups as described in Appendix~\ref{subapp:rsdcorr} (i.e., after velocity correction),\footnote{In principle, one could use the velocity-corrected positions of the Y07 groups (see Appendix~\ref{subapp:rsdcorr}) as `real-space' positions and use these in measuring the 2pcf. This has two disadvantages. Firstly, our current implementation of velocity correction is not exactly the same as in W12, which can lead to systematic errors in the positions. Secondly, the velocity correction only accounts for large scale bulk flows, meaning that we cannot use any satellites in our analysis. For these reasons, we prefer to use the uncorrected (i.e., redshift-space) group positions to calculate all correlation functions. The assignment of $\alpha$ to each group, however, uses the velocity correction as described in Appendix~\ref{subapp:rsdcorr}.} while the mock groups were assigned the value of $\alpha_{2\Mpch}$ of their respective parent halo using measurements from \citet{phs18}. 

\begin{figure*}
\centering
\includegraphics[width=0.9\textwidth]{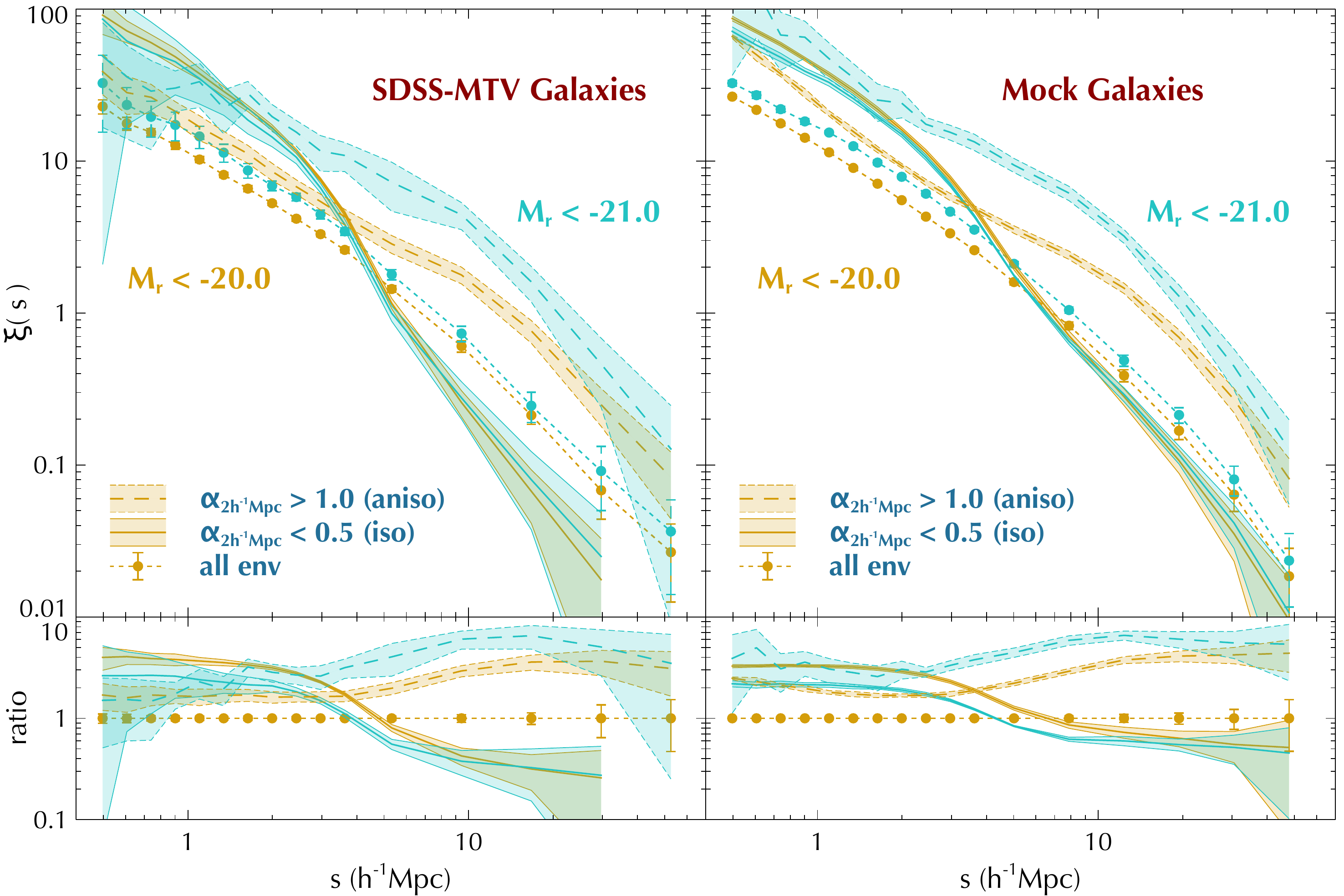}
\caption{\emph{(Top panels:)} 2-point correlation function (2pcf) of group galaxies (lines with error bands) in the SDSS-MTV volume \emph{(top left panel)} and in our mock catalogs \emph{(top right panel)}, selected to be volume-complete for $M_r<-20.0$.
We show the 2pcf for two luminosity thresholds $M_r<-20$ (yellow) and $M_r<-21$ (cyan) for galaxies in anisotropic ($\alpha>1.0$, dashed lines) and isotropic environments ($\alpha<0.5$, solid lines). We see that the luminosity dependence of the clustering is particularly strong in anisotropic environments and is \emph{essentially absent} in isotropic environments. For comparison, the respective all-galaxy measurements from the left panel of Figure~\ref{fig:2pcf-gal-LumCol} are reproduced as the points with error bars in each panel. 
\emph{(Bottom panels:)} Ratio of 2pcf in different environments for the two luminosity thresholds with the all-environment mean trend for the respective threshold. While broadly consistent with the data, the mocks tend to underestimate the relative strength in isotropic environments for each luminosity threshold. See text for a discussion.}
\label{fig:2pcf-gal-Lum-tidal}
\end{figure*}

There is a factor $\sim25$-$30$ difference between the large-scale 2pcf of $\alpha<0.4$ and $\alpha>1.6$ groups, or a factor $\sim 5$ in bias. There are $\sim20,000$ groups with $\alpha<0.4$ and $\sim8000$ with $\alpha>1.6$ in the SDSS-MTV volume. The behaviour with $\alpha$ between $1$-$4\Mpch$ is highly non-monotonic, with all the subsample auto-correlations lying above the auto-correlation of the full sample. Essentially the same effect was also seen by \citet{as07} when splitting SDSS galaxies by large-scale ($\sim8\Mpch$) density contrast. A simple explanation given by those authors is that, at scales smaller than the one used to define environment ($\sim2\Mpch$ in our case), the cross-correlation $\xi_{E_1E_2}$ between objects in two different environments $E_1$ and $E_2$ will satisfy $1+\xi_{E_1E_2}\simeq0$ by construction. It is then easy to show that the auto-correlations $\xi_{E_1E_1}$ and $\xi_{E_2E_2}$ can individually be larger than the auto-correlation of the full sample. It is also easy to see that the precise definition of the environment \citep[tidal anisotropy in our case and overdensity in the case of][]{as07} is irrelevant for the argument.

The points with errors in the \emph{bottom panels} show the ratios of the measurements in different $\alpha$ bins with the mean trend of the all-group measurement, for SDSS-MTV (left) and in the mocks (right). For a better comparison, the mean trends of the ratios in the mocks in the right panel are repeated in the left panel as dashed curves.
The results for mock groups in regions of intermediate tidal anisotropy ($0.4<\alpha<1.6$) agree quite well with the corresponding trends seen in the data.
The clustering of groups in the most anisotropic ($\alpha>1.6$) mock environments, however, is stronger than the corresponding signal in the data by a factor $\sim 2$ at nearly all scales. Similarly, the signal in the most isotropic ($\alpha<0.4$) mock environments is stronger than that in the data at scales $s\gtrsim4\Mpch$, although the jackknife errors for this SDSS subsample become quite large at $s\gtrsim10\Mpch$. We discuss potential reasons for these discrepancies in Section~\ref{sec:discuss}.

Figure~\ref{fig:2pcf-gal-alpha} shows the redshift-space 2pcf of all group galaxies with $M_r<-20.0$, split by tidal anisotropy $\alpha$ as well as colour, in the SDSS-MTV volume \emph{(top left panel)} and in our mock catalogs \emph{(top right panel)}. \emph{Each galaxy was assigned the tidal anisotropy of its group.} The clustering strength is clearly a far stronger function of tidal anisotropy than it is of colour (c.f. also Figure~\ref{fig:2pcf-gal-LumCol}). The 2pcf of the respective full sample is shown for comparison as the points with errors in each panel. The bottom panels show the ratio of the 2pcf of red and blue galaxies in different environments to the mean 2pcf of the full sample. While the overall trends in the mocks are similar to those in the data, we see that the mocks tend to underestimate the difference between the clustering of red and blue galaxies in both isotropic as well as anisotropic environments, similar to the effect we noted for the all-environment measurements in Figure~\ref{fig:2pcf-gal-LumCol}.

Figure~\ref{fig:2pcf-gal-Lum-tidal} shows the redshift-space 2pcf of group galaxies for two luminosity thresholds, restricted to anisotropic (dashed lines) and isotropic environments (solid lines), in the SDSS-MTV volume \emph{(top left panel)} and in our mock catalogs \emph{(top right panel)}. We see that the luminosity dependence of the clustering is particularly strong in anisotropic environments and is essentially absent in isotropic environments. For comparison, the all-galaxy measurements from  Figure~\ref{fig:2pcf-gal-LumCol} are reproduced as the points with error bars in each panel. The \emph{bottom panels} show the ratio of the 2pcf in different environments for the two luminosity thresholds with the all-environment mean trend for the respective threshold. Since the mean trends in the two thresholds are different, this way of presenting the results highlights the relative effect of the environment in a cleaner, more halo-mass-independent manner. We see that the mocks again reproduce the broad trends in the data, but tend to underestimate the relative effect in isotropic environments for each luminosity threshold. 

\begin{figure*}
\centering
\includegraphics[width=0.9\textwidth]{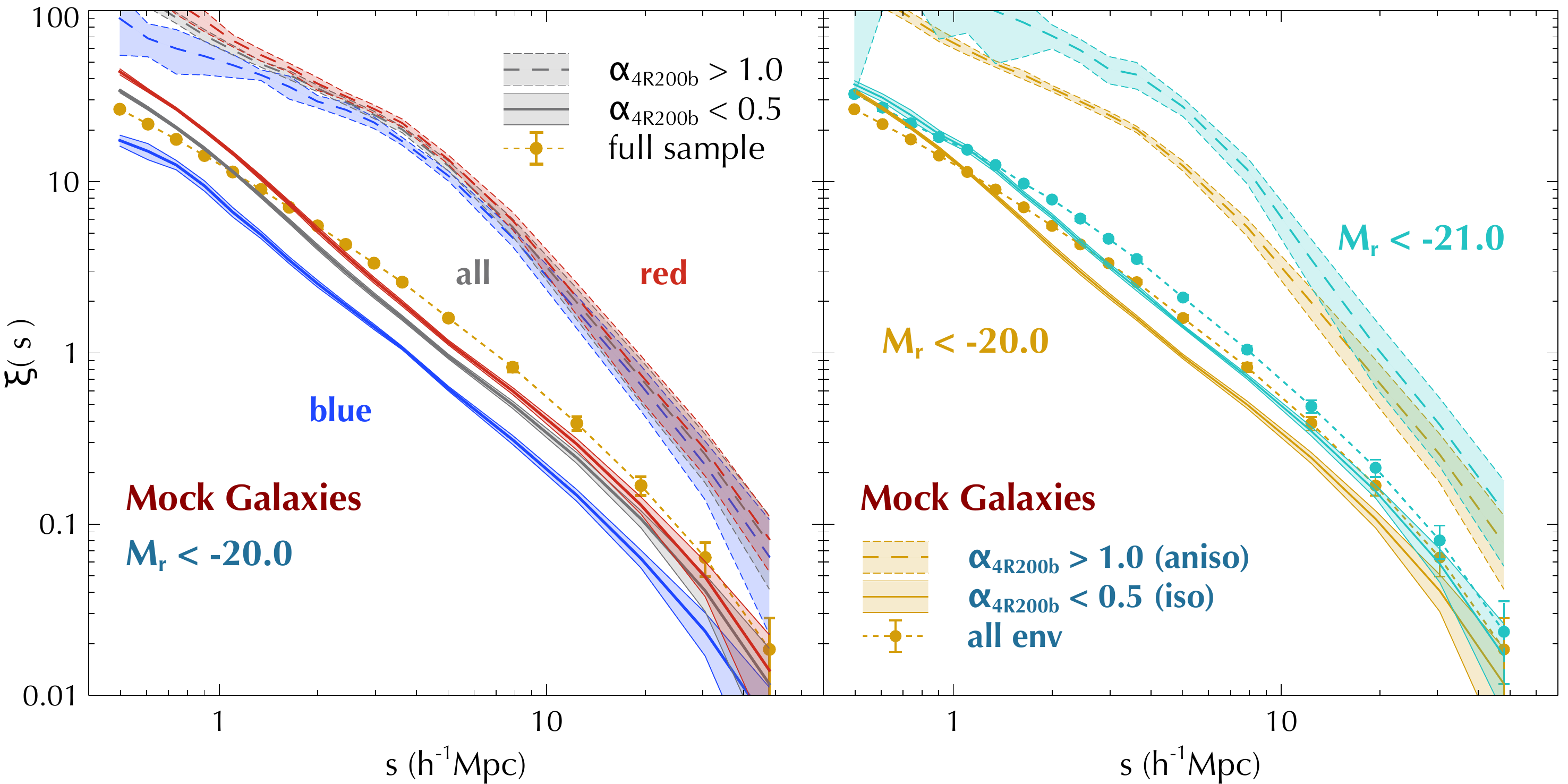}
\caption{Clustering results in mock catalogs split by $\alpha$ defined using a mass-dependent Gaussian smoothing filter, with the filter radius set to the Gaussian equivalent of $4R_{\rm 200b}$. \emph{(Left panel:)} Similar to top right panel of Figure~\ref{fig:2pcf-gal-alpha}. \emph{(Right panel:)} Similar to top right panel of Figure~\ref{fig:2pcf-gal-Lum-tidal}.}
\label{fig:2pcf-mockgal-tidal4R200b}
\end{figure*}

It is also interesting to ask how the results might change if we were to split the tidal environment according to $\alpha$ defined using a mass-dependent Gaussian smoothing scale, with smoothing radius set to the Gaussian equivalent of $4R_{\rm 200b}$ as advocated by \citet{phs18}. While this variable is not available for the observed sample, we do have access to these measurements in the simulations. Figure~\ref{fig:2pcf-mockgal-tidal4R200b} shows the results; we see that using a mass-dependent smoothing leads to a dramatically different luminosity-dependence of clustering in different tidal environments as compared to the $2\Mpch$ case. In isotropic environments, e.g., we now see a luminosity dependence comparable to that in other environments (as well as a large decrease in the overall amplitude of the 2pcf at small scales). 

Following \citet{ss18}, the change in the luminosity-dependence in isotropic environments between Figures~\ref{fig:2pcf-gal-Lum-tidal} and~\ref{fig:2pcf-mockgal-tidal4R200b} can be understood by comparing the relevant values of the smoothing scales involved. \citet{ss18} showed, using overdensity to define environment, that halo mass becomes essentially irrelevant for clustering if the environment defined at a substantially larger scale is held fixed \citep[see also][]{phjg17}. Assuming that similar reasoning is valid for isotropic environments defined using $\alpha$, our results would be explained if the typical halo masses for both luminosity thresholds lead to Gaussian effective smoothing scales substantially less than $2\Mpch$. The HOD used for our mocks leads to typical halo masses of $\log(m_{\rm 200b}/\Mh)\sim12.0\,(12.7)$ for galaxies with $M_r<-20\,(-21)$, corresponding to $R_{\rm 200b}\sim0.25\,(0.4)\Mpch$. The corresponding Gaussian smoothing scales used by \citet{phs18} were $R_{\rm G}=4R_{\rm 200b}/\sqrt{5}\sim0.5\,(0.8)\Mpch$, respectively, both of which are indeed smaller than $2\Mpch$. Refining this model so as to be applicable for all environments is the subject of work in progress.

\section{Discussion}
\label{sec:discuss}
\noindent
We have seen that the (redshift-space) clustering of groups and group galaxies in the SDSS shows strong trends with tidal environment (Figures~\ref{fig:2pcf-alpha}, \ref{fig:2pcf-gal-alpha} and~\ref{fig:2pcf-gal-Lum-tidal}). The clustering strength in each case increases from isotropic to anisotropic environments, as expected from theoretical studies of the environments of dark matter haloes \citep{hahn+09,bprg17,phs18}. Since the simplest statistical models of galaxy properties assume that halo mass is the primary driver of all environmental trends, it is then very interesting to confront these models with the tidal dependence of galaxy clustering in the data.

We have done this by comparing the SDSS results to corresponding measurements in mock catalogs that assumed a `halo mass only' HOD (see section~\ref{subsec:mocks}). On the whole, we see from Figures~\ref{fig:2pcf-alpha}, \ref{fig:2pcf-gal-alpha} and~\ref{fig:2pcf-gal-Lum-tidal} that the mocks match the data in most qualitative aspects, correctly tracking both the luminosity- and colour-dependence of clustering in different tidal environments. There are some notable quantitative differences, however: 
\begin{enumerate}
\item The mocks substantially overestimate the clustering of groups in the most anisotropic and most isotropic environments, at nearly all scales in the former case and at $s\gtrsim4\Mpch$ in the latter (Figure~\ref{fig:2pcf-alpha}).
\item The difference between clustering strengths of red and blue galaxies are somewhat underestimated by the mocks in both isotropic and anisotropic environments (Figure~\ref{fig:2pcf-gal-alpha}) and consequently also in the full sample (Figure~\ref{fig:2pcf-gal-LumCol}).
\item The galaxy clustering strength in isotropic environments relative to all-environment clustering is substantially underestimated in the mocks (bottom panels of Figure~\ref{fig:2pcf-gal-Lum-tidal}).
\end{enumerate}

It is tempting to declare that these differences are evidence that the tidal environment must play a role in galaxy formation and evolution beyond what is captured by the `halo mass only' prescription used in the mocks. However, there are also several technical differences between the mocks and the data which could conceivably be more relevant in explaining the discrepancies noted above. We list the most important of these below.

\begin{itemize}
\item The HOD calibration that forms the basis of our mocks is expected to have systematic errors due to various choices made by \citet{zehavi+11} in their analytical modelling. (E.g., all of the fits for $M_r<-20$ in their Table~3 have reduced Chi-squared values between $\sim2$-$3$.) These fits are being revisited in more recent work \citep[see, e.g.,][]{guo+15}, although we do not yet have access to the corresponding HOD interpolations needed for our mocks.
\item Ideally, we should compare our mock results with the data after applying the Y07 algorithm to the mocks. This could be important since the Y07 algorithm introduces spurious effects due to central/satellite misclassifications as well as member misallocations \citep{campbell+15}. It is very likely that these effects have different strengths in isotropic and anisotropic tidal environments. As it stands, these effects are present in the data but not in our mocks.
\item The halo mass assignment in the Y07 algorithm is known to have a scatter of $\sim0.3$dex \citep{yang+07,campbell+15}. This will affect objects near the selection thresholds of our luminosity- and halo mass-selected samples, and could also therefore affect objects systematically as a function of tidal environment (since, e.g., anisotropic environments preferentially host low mass haloes).
\end{itemize}

Appendix~\ref{subapp:mockVsdata} lists some additional technical differences between the data and our mocks, which however are expected to play a much smaller role than the ones discussed above. Given the uncertainties discussed above, it is difficult to attribute all of the differences seen between the mocks and the data to `beyond halo mass' effects that are missing in the mocks. We therefore cautiously conclude that a `halo mass only' HOD prescription seems to describe the basic effects of tidal environment on luminosity- and colour-dependent clustering quite well -- i.e., galaxy evolution variables apparently do not need to explicitly depend on tidal environment at this level of comparison. 

We do note, however, that the rather large differences seen between mocks and data in group clustering in the most and least anisotropic environments (Figure~\ref{fig:2pcf-alpha}), as well as the differences in relative strengths of galaxy clustering in isotropic environments to the all-environment clustering (bottom panels of Figure~\ref{fig:2pcf-gal-Lum-tidal}), deserve further attention. 
In forthcoming work, we will also explore the behaviour of galaxy red fractions and the nature of galactic conformity\footnote{Galactic conformity is the effect wherein satellite galaxies `know' about the star formation properties of their host central galaxy, with star forming centrals preferentially hosting star forming satellites \citep{weinmann+06}. It is interesting to ask whether the level of conformity depends on tidal environment; this could have implications for both HOD analyses as well as semi-analytical models of galaxy formation.} in different tidal environments.

\section{Conclusions}
\label{sec:conclude}
\noindent
We have explored the dependence of galaxy clustering in the SDSS on local tidal environment. We have focused on the redshift-space clustering of groups and group galaxies, using the group catalog of \citet[][Y07]{yang+07}. For the tidal environment, we have used measurements of the tidal tensor eigenvalue field provided by \citet[][W12]{wmyv12} and assigned values of tidal anisotropy $\alpha$ \citep[equation~\ref{eq:alphadef}; for details, see][]{phs18} to groups and their member galaxies. As expected from previous theoretical studies, we see a strong dependence of clustering strength on tidal environment, with the strength increasing from isotropic (node-like) to anisotropic (filament-like) environments.

We have also attempted to assess whether or not the tidal environment leaves a direct imprint on galaxy formation and evolution, beyond what is already captured by a dependence on halo mass. To this end, we have compared the clustering measurements in the data with those in mock catalogs generated using a `mass only HOD'. \emph{The mocks qualitatively reproduce all the trends seen in the data, although there are several quantitative differences.} Many of these differences, however, could be attributable to technical differences between our mocks and the data, so that galaxy evolution variables likely do not need to depend explicitly on tidal environment. 

However, there remain some intriguingly large differences between the clustering of groups and of galaxies in isotropic environments in mocks and data (see Figures~\ref{fig:2pcf-alpha} and~\ref{fig:2pcf-gal-Lum-tidal} and the discussion in section~\ref{sec:discuss}) which we believe deserve further attention. We will return to this comparison in future work, along with comparisons between the SDSS tidal environment inferred by W12 and that using other techniques such as the Bayesian inference BORG algorithm of \citet{jlw15}.

As a final remark, we note from Figure~\ref{fig:2pcf-gal-alpha} that factors of $\sim20$ in 2pcf ratios are easily achievable using bright galaxies with number densities of $\sim6\times10^{-3}(\Mpch)^{-3}$. Negative ratios could also be possible using cross-correlations. This should be interesting for multi-tracer studies that seek to maximise the difference in clustering for co-located samples \citep[see, e.g.,][]{mCds09,hsd11,fcsm15}.  To date, such studies have been confined to samples selected as in Figure~\ref{fig:2pcf-gal-LumCol}, which only show factor of $\sim 2$ differences.

{\bf Note added:} While this work was being completed, we became aware of a similar analysis by \citet{azpm18}. These authors have analysed the tidal environment of SDSS galaxies using a different set of selection criteria, analysis techniques and definition of tidal anisotropy. Encouragingly, the results of our two analyses, wherever they can be compared, are qualitatively consistent.

\section*{Acknowledgments}
We are indebted to the authors of \citet{yang+07} and \citet{wmyv12} for generously making their data sets publicly available in user-friendly formats. We also acknowledge the SDSS team, whose data forms the basis of all these analyses. AP gratefully acknowledges use of high performance computing facilities at IUCAA, Pune (http://hpc.iucaa.in). The research of AP is supported by the Associateship Scheme of ICTP, Trieste and the Ramanujan Fellowship awarded by the Department of Science and Technology, Government of India. OH acknowledges funding from the European Research Council (ERC) under the European Union's Horizon 2020 research and innovation programme (grant agreement No. 679145, project `COSMO-SIMS'). We thank Shadab Alam for useful discussions and correspondence.

\bibliography{masterRef}

\appendix

\section{Technical notes}
\label{app:technical}

\subsection{$\alpha$-assignment: correcting group redshifts for peculiar velocity effects}
\label{subapp:rsdcorr}
\noindent
To assign a value of tidal anisotropy $\alpha$ to each group, we need the group's `real-space' position. We estimate this directly from the velocity field provided by W12 as follows. 

The data set gives the velocity field smoothed with a Gaussian kernel enclosing $10^{13}\Mh$ ($R\sim2\Mpch$). W12 performed a peculiar velocity correction at a smoothing of $10^{14.75}\Mh$ (corresponding to a Gaussian with scale $R_{\rm s}=7.93\Mpch$, see eqn 8 of W12). To match this, we first smooth the velocity data with a Gaussian kernel of scale $R_{\rm eff} = R_{\rm s}\times\sqrt{1-10^{(-1.75/3)\times2}}=0.9653 R_{\rm s}\simeq7.66\Mpch$. Each group is then assigned the velocity of its enclosing grid cell, which is then used to calculate the group's line-of-sight velocity component $v_{\rm los} = \vec{v}\cdot\hat r$ where $\hat r$ is the unit vector along the group position (which is independent of the group redshift and can hence be calculated as $\hat r = \vec{r}(z_{\rm obs})/r(z_{\rm obs})$ using the observed redshift $z_{\rm obs}$). Following W12, the corrected redshift $z_{\rm corr}$ is then given by $z_{\rm corr} = (z_{\rm obs} - v_{\rm los}/c)/(1+v_{\rm los}/c)$. We find that this procedure leads to a distribution of $v_{\rm los}/c$ with mean $\sim-3\times10^{-4}$ and standard deviation $\sim6\times10^{-4}$ (corresponding to distance corrections $\sim(-0.9\pm1.8)\Mpch$ at $z=0$). 

We then assign each group the $\alpha$ value of the grid cell enclosing the group's corrected position. \emph{We do not use the corrected position any further.} In other words, all our correlation functions are measured in redshift-space. This procedure ensures that uncertainties due to the velocity correction procedure are minimised. (Recall that $\alpha$ is defined using a Gaussian filtering scale of $\sim2\Mpch$, or a TopHat equivalent scale of $\sim4.5\Mpch$, which is substantially larger than the typical velocity correction.)

\subsection{Additional differences between mocks and data}
\label{subapp:mockVsdata}
\noindent
Here we list some technical differences between the SDSS data and our mocks which might have a bearing on the comparison of environmental trends in each. The differences listed below are expected to play a much smaller role than those listed in section~\ref{sec:discuss}.

\begin{itemize}
\item The value of $\Omega_{\rm m}$ in the simulations used for our mocks is $\sim10\%$ larger than that used for the HOD. It is unlikely, however, that this would lead to environment-dependent differences, especially at the level seen in Figure~\ref{fig:2pcf-alpha}, for example.
\item Our mocks do not have position/velocity offsets between group galaxies and their respective parent haloes. These effects, however, are expected to be much smaller than the differences we have noted in the main text \citep[see, e.g., Figure 3 of][]{guo+15b}.
\item When measuring the tidal field, W12 used Gaussian smoothing with a radius enclosing $10^{13}\Mh$ which translates to $R\simeq2.07\Mpch$ (see equation~8 of W12), while we use measurements at exactly $R=2\Mpch$ in our simulations.
\item As noted by W12, their measurements of the tidal tensor eigenvalues $\lambda_1,\lambda_2,\lambda_3$ at $R\sim2\Mpch$ are affected by a fairly large scatter, as well as a substantial systematic offset for $\lambda_2$ and $\lambda_3$ (see their Figure 5, top panel). W12 attribute this mostly to inaccuracies inherent in their method and in the Y07 group finder. While these errors propagate into the values of $\alpha$ used by us through \eqn{eq:alphadef}, systematic errors would tend to partially cancel since $\alpha$ involves a ratio of eigenvalues.
\item Our assignment of $\alpha$ to the Y07 groups is affected by the systematic uncertainties discussed in section~\ref{subsec:sdss} and Appendix~\ref{subapp:rsdcorr}.
\end{itemize}

\label{lastpage}

\end{document}